\documentstyle[aps,prl,epsf,twocolumn]{revtex}
\begin{document}
\def\B.#1{{\bbox{#1}}}
\draft
\title{{Hydrodynamic Turbulence: a 19th Century Problem 
\\ with a challenge
for the 21st Century}}
\author { Victor L'vov  and Itamar Procaccia}
\address{Department of~~Chemical Physics,
The Weizmann Institute of Science,
Rehovot 76100, Israel}
\maketitle
\vskip 1cm

\section{Turbulence for the Physicist and for the Engineer}

Sir Horace Lamb once said ``I am an old man now, and when I die and go
to Heaven there are two matters on which I hope enlightenment. One is
quantum electro-dynamics and the other is turbulence of fluids. About
the former, I am really rather optimistic"\cite{69Gol}.  Possibly
Lamb's pessimism about turbulence had been short-sighted. There exist
signs that the two issues that concerned Lamb are not disconnected.
The connections were brewing for some while, and began to take clearer
form recently. This promises renewed vigor in the intellectual pursuit
for understanding this long-standing problem. This is not due to some
outstanding developments of new tools in the theoretical or
experimental study of turbulence {\it per se}, but rather due to
developments in neighboring fields. The great successes of the theory
of critical and chaotic phenomena and the popularity of nonlinear
physics of classical systems attracted efforts that combined the
strength of fields like quantum field theory and condensed matter
physics leading to renewed optimism about the solubility of the
problem of turbulence. It seems that this area of research will have a
renaissance of rapid growth that promises excitement well into the
next century.

It is possibly a happy coincidence that hydrodynamic turbulence is
considered a problem of immense interest by both physicists and
engineers. The physicist tends to appreciate phenomena that display
universal characteristics; the engineer may find such characteristics
irrelevant since they cannot be manipulated. The engineer seeks
control, and control means a ready response to perturbations.
Universal phenomena are immune to perturbations. The point is of
course that ``turbulence" means different things to different
researchers. All agree that hydrodynamic turbulence arises in fluids
that are highly stressed, or stirred, such that there exist
significant fluid velocities (or winds) on the largest scales of
motion. The engineer is typically interested in the flow
characteristics near the boundaries of the fluid (boundary layers,
airplane wings, pipes, turbines etc). By understanding how to
manipulate the boundary region one may reduce drag and improve the
performance of technological devices. The physicist is interested in
the small scale structure of turbulence away from any boundary, where
the action of fluid mechanics effectively homogenizes the flow
characteristics and where universal phenomena may be sought. In this
context ``universal" are those phenomena that are independent of the
nature of the fluid (water, oil, honey etc), independent of the
mechanism of stirring the flow, and independent of the form of the
container of the fluid. They are inherent features of fluid mechanics
as a classical field theory.  Understanding these universal features
can gain only marginal technological improvements. But this is the
theoretical challenge that excites the physicist.

To many people turbulence research seems orthogonal to the two main
lines of progress in modern physics. On the one hand, tremendous
effort has been invested in understanding the structure of matter,
with later developments concentrating on ever-diminishing scales of
constituent particles using the ever-increasing energies of particle
accelerators.  On the other hand astronomy and cosmology have exploded
with a rich tableau of discoveries at ever-increasing distances from
our galaxy.  The physics of phenomena on the human scale, phenomena
that are of acute interest to the scientist and layman alike, were
relegated into a secondary position in the course of the development
of the first half of 20th century physics.  Of course, problems
related to the health and well-being of humans are deservedly being
studied in biology and medicine. But {\rm physical} phenomena that can
be observed by simply looking out the window are considered by many as
``non-fundamental" and belonging to 19th century research. It is the
conviction of the present writers as well as of a growing number of
researchers that physics on the human scale offers tremendously
rewarding intellectual challenges, some of which were at the core of
the recent interest in chaotic phenomena and in the area which is
vaguely termed ``physics of complex systems". Fluid turbulence, which
is the highly complex, chaotic and vortical flow that is
characteristic of all fluids under large stresses, is a paramount
example of these phenomena that are immensely challenging to the
physicist and the mathematician alike. The aim of this paper is to
explain why this problem is exciting, why it is difficult, and what
are the possible routes that one can traverse in finding the solution.
The point of view described here is that of the physicist whose
interest is biased in favour of universal phenomena.

\section{Some History}

The mathematical history of fluid mechanics begins with Leonhard Euler
who was invited by Frederick the Great to Potsdam in 1741. According
to a popular story (which we could not corroborate) one of his tasks
was to engineer a water fountain.  As a true theorist, he began by
trying to understand the laws of motion of fluids. In 1755 he wrote
Newton's laws for a fluid which in modern notation density reads (for
the case of constant density) \cite{Lam}
\begin{equation}
{\partial \B. u (\B.r,t) \over \partial t} + \B.u (\B.r,t)\cdot \B.\nabla \B.u
(\B.r,t) = -\B.\nabla p(\B.r,t) \ . \label{euler}
\end{equation}
Here $\B.u(\B.r,t)$ and $p(\B.r,t)$ are the fluid velocity and
pressure at the spatial point $\B.r$ at time $t$.  The LHS of this
``Euler equation" for ${\bf u}({\bf r},t)$ is just the material time
derivative of the momentum, and the RHS is the force, which is
represented as the gradient of the pressure imposed on the fluid. In
fact, trying to build a fountain on the basis of this equation was
bound to fail. This equation predicts, for a given gradient of
pressure, velocities that are much higher than anything observed. One
missing idea was that of the viscous dissipation that is due to the
friction of one parcel of fluid against neighboring ones. The
appropriate term was added to (\ref{euler}) by Navier in 1827 by
Stokes in 1845\cite{Lam}. The result is known as the ``Navier-Stokes
equations" :
\begin{equation}
{\partial \B.u (\B.r,t) \over \partial t} +\B.u (\B.r,t)\cdot \B.\nabla \B.u
(\B.r,t) = -\B.\nabla p(\B.r,t) +\nu \nabla^2\B.u (\B.r,t)
\ . \label{NS}
\end{equation}
Here $\nu$ is the kinematic viscosity, which is about $10^{-2}$ and
$0.15~{\rm cm^2/sec}$ for water and air at room temperature
respectively. Without the term $\nu\nabla^2\B.u (\B.r,t)$ the kinetic
energy $u^2/2$ is conserved; with this term kinetic energy is
dissipated and turned into heat. The effect of this term is to
stabilize and control the nonlinear energy conserving Euler equation
(\ref{euler}).

Straightforward attempts to assess the solutions of this equation may
still be very non-realistic.  For example, we could estimate the
velocity of water flow in any one of the mighty rivers like the Nile
or the Volga which drop hundreds of meters in a course of about a
thousand kilometers. The typical angle of inclination $\alpha$ is
about $10^{-4}$ radians, and the typical river depth $L$ is about 10
meters. Equating the gravity force $\alpha g$ ($g\simeq 10^3 {\rm
  cm/sec^2}$) and the viscous drag $\nu d^2 u/dz^2 \sim \nu u/L^2$ we
find $u$ to be of the order of $10^7$ cm/sec instead of the observed
value of about $10^2$ cm/sec.  This is of course absurd, perhaps to
the regret of the white water rafting industry.  The resolution of
this discrepancy was suggested by Reynolds \cite{Rey} who stressed the
importance of a dimensionless ratio of the nonlinear term to the
viscous term in (\ref{NS}). With a velocity drop of the order of $U$
on a scale $L$ the nonlinear term is estimated as $U^2/L$.  The
viscous term is about $\nu U/L^2$. The ratio of the two, known as the
Reynolds number Re, is $UL/\nu$. The magnitude of Re measures how
large is the nonlinearity compared to the effect of the viscous
dissipation in a particular fluid flow.  For Re$\ll 1$ one can neglect
the nonlinearity and the solutions of the Navier-Stokes equations can
be found in close-form in many instances \cite{Lam}. In natural
circumstances Re is very large. For example in the rivers discussed
above Re$\simeq 10^7$. Reynolds understood that for ${\rm Re}\gg 1$
there is no stable stationary solution for the equations of motion.
The solutions are strongly affected by the nonlinearity, and the
actual flow pattern is complicated, convoluted and vortical.  Such
flows are called turbulent.

Modern concepts about high Re number turbulence started to evolve with
Richardson's insightful contributions \cite{Ric} which contained the
famous ``poem" that paraphrased J. Swift: "{\it Big whirls have little
  whirls that feed on their velocity, and little whirls have lesser
  whirls and so on to viscosity -in the molecular sense}". In this way
Richardson conveyed an image of the creation of turbulence by large
scale forcing, setting up a cascade of energy transfers to smaller and
smaller scales by the nonlinearities of fluid motion, until the energy
dissipates at small scales by viscosity, turning into heat. This
picture led in time to innumerable ``cascade models" that tried to
capture the statistical physics of turbulence by assuming some thing
or other about the cascade process. Indeed, no one in their right mind
is interested in the full solution of the turbulent velocity field at
all points in space-time. The interest is in the statistical
properties of the turbulent flow. Moreover the statistics of the
velocity field itself is too heavily dependent on the particular
boundary conditions of the flow. Richardson understood that universal
properties may be found in the statistics of velocity {\em
  differences} $\delta \B.u( \B.r_1,\B.r_2)\equiv
\B.u(\B.r_2)-\B.u(\B.r_1)$ across a separation $\B.R=\B.r_2-\B.r_1$.
In taking such a difference we subtract the non-universal large scale
motions (known as the ``wind" in atmospheric flows). In experiments
(see for example \cite{MY-2,84AGHA,93SK,93Ben,Fri}) it is common to
consider one dimensional cuts of the velocity field, $\delta
u_{\ell}(R)\equiv \delta\B.u(\B.r_1,\B.r_2)\cdot\B.R/R$.  The interest
is in the probability distribution function of $\delta u_{\ell}(R)$
and its moments. These moments are known as the ``structure functions"
$S_n(R)\equiv \left\langle\delta u_{\ell}(R)^n\right\rangle$ where
$\left<\dots\right>$ stands for a suitably defined ensemble average.
For Gaussian statistics the whole distribution function is governed by
the second moment $S_2(R)$, and there is no information to be gained
from higher order moments. In contrast, hydrodynamic experiments
indicate that turbulent statistics is extremely non-Gaussian, and the
higher order moments contain important new information about the
distribution functions.

Possibly the most ingenious attempt to understand the statistics of
turbulence is due to Kolmogorov who in 1941 \cite{41Kol} proposed the
idea of universality (turning the study of small scale turbulence from
mechanics to fundamental physics) based on the notion of the
``inertial range". The idea is that for very large values of Re there
is a wide separation between the ``scale of energy input" $L$ and the
typical ``viscous dissipation scale" $\eta$ at which viscous friction
become important and dumps the energy into heat. In the stationary
situation, when the statistical characteristics of the turbulent flow
are time independent, the rate of energy input at large scales ($L$)
is balanced by the rate of energy dissipation at small scales
($\eta$), and must be also the same as the flux of energy from larger
to smaller scales (denoted $\bar \epsilon$) as it is measured at any
scale $R$ in the so-called ``inertial" interval $\eta\ll R \ll L$.
Kolmogorov proposed that the only relevant parameter in the inertial
interval is $\bar \epsilon$, and that $L$ and $\eta$ are irrelevant
for the statistical characteristics of motions on the scale of $R$.
This assumption means that $R$ is the only available length for the
development of dimensional analysis. In addition we have the
dimensional parameters $\bar\epsilon$ and the mass density of the
fluid $\rho$.  From these three parameters we can form combinations
$\rho^x\bar\epsilon^yR^z$ such that with a proper choice of the
exponents $x,y,z$ we form any dimensionality that we want. This leads
to detailed predictions about the statistical physics of turbulence.
For example, to predict $S_n(R)$ we note that the only combination of
$\bar \epsilon$ and $R$ that gives the right dimension for $S_n$ is
$\left (\bar \epsilon R \right )^{n/3}$. In particular for $n=2$ this
is the famous Kolmogorov ``2/3" law which in Fourier representation is
also known as the ``-5/3" law.  The idea that one extracts universal
properties by focusing on statistical quantities can be applied also
to the correlations of gradients of the velocity field.  An important
example is the rate $\epsilon({\bf r},t)$ at which energy is
dissipated into heat due to viscous damping. This rate is roughly $
\nu|\nabla {\bf u}({\bf r},t)|^2$. One is interested in the
fluctuations of the energy dissipation $\epsilon(\B.r,t)$ about their
mean $\bar\epsilon$, $\hat\epsilon({\bf r},t) = \epsilon ({\bf r},t) -
\bar\epsilon$, and how these fluctuations are correlated in space. The
answer is given by the often-studied correlation function
$K_{\epsilon\epsilon}(R) = \left< \hat\epsilon ({\bf r}+{\bf R},t)
  \hat\epsilon ({\bf r},t) \right>$. If the fluctuations at different
points were uncorrelated, this function would vanish for all $R\ne 0$.
Within the Kolmogorov theory one estimates $K_{\epsilon\epsilon}(R)
\simeq \nu^2\bar \epsilon^{4/3}R^{-8/3}$, which means that the
correlation decays as a power, like $1/R^{8/3}$.

Experimental measurements show that Kolmogorov was remarkably close to
the truth.  The major aspect of his predictions, i.e. that the
statistical quantities depend on the length scale $R$ as power laws is
corroborated by experiments. On the other hand, the predicted
exponents seem not to be exactly realized.  For example, the
experimental correlation $K_{\epsilon\epsilon}(R)$ decays according to
a power law, $K_{\epsilon\epsilon}(R) \sim R^{-\mu}$ for $      \eta
\ll R \ll L$, with $\mu$ having a numerical value of $0.2-0.3$ instead
of 8/3 \cite{93SK}.  The structure functions also behave as power
laws, $S_n(R) \simeq R^{\zeta_n}$, but the numerical values of
$\zeta_n$ deviate progressively from $n/3$ when $n$ increases
\cite{84AGHA,93Ben}.  Something fundamental seems to be missing. The
uninitiated reader might think that the numerical value of this
exponent or another is not a fundamental issue.  However one needs to
understand that the Kolmogorov theory exhausts the dimensions of the
statistical quantities under the assumption that $\bar\epsilon$ is the
only relevant parameter. Therefore a deviation in the numerical value
of an exponent from the prediction of dimensional analysis requires
the appearance of another dimensional parameter. Of course there
exists two dimensional parameters, i.e. $L$ and $\eta$ which may turn
out to be relevant.  Indeed, experiments indicated that for the
statistical quantities mentioned above the energy-input scale $L$ is
indeed relevant and it appears as a normalization scale for the
deviations from Kolmogorov's predictions: $S_n(R)\simeq \left (\bar
  \epsilon R \right )^{n/3}\left(L/R\right)^{\delta_n}$ where
$\zeta_n=n/3-\delta_n$.  Such form of scaling which deviates from the
predictions of dimensional analysis is referred to as ``anomalous
scaling".  The realization that the experimental results for the
structure functions were consistent with $L$ rather than $\eta$ as the
normalization scale developed over a long time and involved a large
number of experiments; recently the accuracy of determination of the
exponents increased appreciably as a result of a clever method of data
analysis by Benzi, Ciliberto and coworkers\cite{93Ben}.  Similarly a
careful demonstration of the appearance of $L$ in the dissipation
correlation was achieved by Sreenivasan and coworkers \cite{93SK}. A
direct analysis of scaling exponents $\zeta_n$ and $\mu$ in a high
Reynolds number flow was presented by Praskovskii and Oncley, leading
to the same conclusions \cite{95Pra}.

\section{Turbulence as a Field Theory}

Theoretical studies of the universal small scale structure of
turbulence can be classified broadly into two main classes. Firstly
there is a large body of phenomenological models that by attempting to
achieve agreement with experiments reached important insights on the
nature of the cascade or the statistics of the turbulent fields
\cite{Fri}. In particular there appeared influential ideas, following
Mandelbrot \cite{74Man}, about the fractal geometry of highly
turbulent fields which allow scaling properties that are sufficiently
complicated to include also non-Kolmogorov scaling.  Parisi and Frisch
showed that by introducing multifractals one can accomodate the
nonlinear dependence of $\zeta_n$ and $n$ \cite{85PF}. However these
models are not derived on the basis of the equations of fluid
mechanics; one is always left with uncertainties about the validity or
relevance of these models. The second class of approaches is based on
the equations of fluid mechanics. Typically one acknowledges the fact
that fluid mechanics is a (classical) field theory and resorts to
field theoretic methods in order to compute statistical quantities.
Even though there had been a continuous effort during almost 50 years
in this direction, the analytic derivation of the scaling laws for
$K_{\epsilon\epsilon}(R)$ and $S_n(R)$ from the Navier-Stokes
equations (\ref{NS}) and the calculation of the numerical value of the
scaling exponents $\mu$ and $\zeta_n$ have been among the most elusive
goals of theoretical research. Why did it turn out to be so difficult?

To understand the difficulties, we need to elaborate a little on the
nature of the field theoretic approach. Suppose that we want to
calculate the average response of a turbulent fluid at some point
${\bf r}_0$ to forcing at point ${\bf r}_1$.  The field theoretic
approach allows us to consider this response as an infinite sum of all
the following processes: firstly there is the direct response at point
${\bf r}_0$ due to the forcing at ${\bf r}_1$. This response is
instantaneous if we assume that the fluid is incompressible (and
therefore the speed of sound is infinite).  Then there is the process
of forcing at ${\bf r}_1$ with a response at an intermediate point
${\bf r}_2$, which then acts as a forcing for the response at ${\bf
  r}_0$.  This intermediate process can take time, and we need to
integrate over all the possible positions of point ${\bf r}_2$ and all
times. This is the second-order term in perturbation theory. Then we
can force at ${\bf r}_1$, the response at ${\bf r}_2$ acting as a
forcing for ${\bf r}_3$ and the response at ${\bf r}_3$ forces a
response at ${\bf r}_0$. We need to integrate over all possible
intermediate positions ${\bf r}_2$ and ${\bf r}_3$ and all the
intermediate times. This is the third-order term in perturbation
theory. And so on. The actual response is the infinite sum of all
these contributions.  In applying this field theoretical method one
encounters three main difficulties:

(A) The theory has no small parameter. The usual procedure is to
develop the theory perturbatively around the linear part of the
equation of motion.  In other words, the zeroth order solution of
Eq.(\ref{NS}) is obtained by discarding the terms which are quadratic
in the velocity field. The expansion parameter is then obtained from
the ratio of the quadratic to the linear terms; this ratio is of
theorder of Reynolds number Re which was defined above. Since we are
interested in Re$\gg 1$, naive perturbation expansions are badly
divergent. In other words the contribution of the various processes
described above increases as $\left({\rm Re}\right)^n$ with the number
$n$ of intermediate points in space-time.

(B) The theory exhibits two types of nonlinear interactions. Both are
hidden in the nonlinear term $\B.u \cdot \nabla \B.u$ in Eq.
(\ref{NS}). The larger of the two is known to any person who watched
how a small floating object is entrained in the eddies of a river and
swept along a complicated path with the turbulent flow. In a similar
way any fluctuation of small scale is swept along by all the larger
eddies. Physically this sweeping couples any given scale of motion to
all the larger scales. Unfortunately the largest scales contain most
of the energy of the flow; these large scale motions are what is
experienced as gusts of wind in the atmosphere or the swell in the
ocean.  In the perturbation theory for $S_n(R)$ one has the
consequences of the sweeping effect from all the scale larger than
$R$, with the main contribution coming from the largest, most
intensive gusts on the scale of $L$. As a result these contributions
diverge when $L\to\infty$. In the theoretical jargon this is known as
``infrared divergences". Such divergences are common in other field
theories, with the best known example being quantum electrodynamics.
In that theory the divergences are of similar strength in higher order
terms in the series, and they can be removed by introducing finite
constants to the theory, like the charge and the mass of the electron.
In the hydrodynamic theory the divergences become stronger with the
order of the contribution, and to eliminate them in this manner one
needs an infinite number of constants. In the jargon such a theory is
called ``not renormalizable". However, sweeping is just a kinematic
effect that does not lead to energy redistribution between scales, and
one may hope that if the effect of sweeping is taken care of in a
consistent fashion a renormalizable theory might emerge.  This
redistribution of energy results from the second type of interaction,
that stems from the shear and torsion effects that are sizable only if
they couple fluid motions of comparable scales. The second type of
nonlinearity is smaller in size but crucial in consequence, and it may
certainly lead to a scale-invariant theory.

(C) Nonlocality of interaction in ${\bf r}$ space. One recognizes that
the gradient of the pressure is dimensionally the same as
$(\B.u\cdot\bbox{\nabla})\B.u$, and the fluctuations in the pressure
are quadratic in the fluctuations of the velocity. This means that the
pressure term is also nonlinear in the velocity. However, the pressure
at any given point is determined by the velocity field everywhere.
Theoretically one sees this effect by taking the divergence of
Eq.(\ref{NS}). This leads to the equation $\nabla^2 p
=\bbox{\nabla}\cdot[(\B.u\cdot\bbox{\nabla})\B.u]$. The inversion of
the Laplacian operator involves an integral over all space. Physically
this stems from the fact that in the incompressible limit of the
Navier-Stokes equations sound speed is infinite and velocity
fluctuations in all distant points are instantaneously coupled.

Indeed, these difficulties seemed to complicate the application of
field theoretic methods to such a degree that a wide-spread feeling
appeared to the effect that it is impossible to gain valuable insight
into the universal properties of turbulence along these lines, even
though they proved so fruitful in other field theories. The present
authors (as well as other researchers starting with Kraichnan \cite{}
and recently Migdal \cite{94Mig}, Polyakov \cite{93Pol}, Eyink
\cite{93Eyi} etc.) think differently, and in the rest of this paper we
will explain why.

The first task of a successful theory of turbulence is to overcome the
existence of the interwoven nonlinear effects that were explained in
difficulty (B). This is not achieved by directly applying a formal
field-theoretical tool to the Navier-Stokes equations. It does not
matter whether one uses standard field theoretic perturbation theory
\cite{61Wyl}, path integral formulation, renormalization group
\cite{86Yak} $\epsilon$-expansion, large $N$-limit \cite{95MW} or
one's formal method of choice.  One needs to take care of the
particular nature of hydrodynamic turbulence as embodied in difficulty
(B) {\it first}, and {\it then} proceed using formal tools.

The removal of the effects of sweeping is based on Richardson's remark
that universality in turbulence is expected for the statistics of
velocity {\it differences across a length scale} $R$ rather than for
the statistics of the velocity field itself. The velocity fields are
dominated by the large scale motions that are not universal since they
are produced directly by the agent that forces the flow. This forcing
agent differs in different flow realizations (atmosphere, wind
tunnels, channel flow etc.).  Richardson's insight was developed by
Kraichnan who attempted to cast the field theoretic approach in terms
of Lagrangian paths, meaning a description of the fluid flow which
follows the paths of every individual fluid particle. Such a
description automatically removes the large scale contributions
\cite{65Kra}.  Kraichnan's approach was fundamentally correct, and
gave rise to important and influential insights in the description of
turbulence, but did not provide a convenient technical way to consider
all the orders of perturbation theory. The theory does not provide
transparent rules how to consider an arbitrarily high term in the
perturbation theory. Only low order truncations were considered.

A way to overcome difficulty (B) was suggested by Belinicher and L'vov
\cite{87BL} who introduced a novel transformation that allowed on one
hand the elimination of the sweeping that leads to infrared
divergences, and on the other hand allows the development of simple
rules for writing down any arbitrary order in the perturbation theory
for the statistical quantities. The essential idea in this
transformation is the use of a coordinate frame in which velocities
are measured relative to the velocity of {\em one} fluid particle.
The use of this transformation allowed the examination of the
structure functions of velocity differences $S_n(R)$ to all orders in
perturbation theory.  Of course, difficulty (A) remains; the
perturbation series still diverges rapidly for large values of Re, but
now standard field theoretic methods can be used to reformulate the
perturbation expansion such that the viscosity is changed by an
effective ``eddie viscosity". The theoretical tool that achieves this
exchange is known in quantum field theory as the Dyson line
resummation \cite{LP-1}. The result of this procedure is that the
effective expansion parameter is no longer Re but an expansion
parameter of the order of unity.  Of course, such a perturbation
series may still diverge as a whole. Notwithstanding it is crucial to
examine first the order-by-order properties of series of this type.

Such an examination leads to a major surprise: every term in this
perturbation theory remains finite when the energy-input scale $L$
goes to $\infty$ and the viscous-dissipation scale $\eta$ goes to $0$
\cite{LP-1}. The meaning of this is that the perturbative theory for
$S_n$ does not indicate the existence of any typical length-scale.
Such a length is needed in order to represent deviations in the
scaling exponents from the predictions of Kolmogorov's dimensional
analysis in which both scales $L$ and $\eta$ are assumed irrelevant.
In other areas of theoretical physics in which anomalous scaling has
been found it is common that already the perturbative series indicates
this phenomenon.  In many cases this is seen in the appearance of
logarithmic divergences that must be tamed by truncating the integrals
at some renormalization length. Hydrodynamic turbulence seems at this
point different. The nonlinear Belinicher-L'vov transformation changes
the underlying linear theory such that the resulting perturbative
scheme for the structure functions is finite order by order
\cite{87BL,LP-1}.  The physical meaning of this result is that as much
as can be seen from this perturbative series the main effects on the
statistical quantities for velocity differences across a scale $R$
come from activities of scales comparable to $R$. This is the
perturbative justification of the Richardson-Kolmogorov cascade
picture in which widely separated scales do not interact.

Consequently the main question still remains: how does a
renormalization scale appear in the statistical theory of turbulence?

It turns out that there are two different mechanisms that furnish a
renormalization scale, and that finally {\em both} $L$ {\em and}
$\eta$ appear in the theory. The viscous scale $\eta$ appears via a
rather standard mechanism that can be seen in perturbation theory as
logarithmic divergences, but in order to see it one needs to consider
the statistics of gradient fields rather than the velocity differences
themselves \cite{LL,LP-2}.  For example, considering the perturbative
series for $K_{\epsilon\epsilon}(R)$, which is the correlation
function of the rate of energy dissipation $\nu|\bbox{\nabla}\B.u|^2$,
leads immediately to the discovery of logarithmic ultraviolet
divergences in every order of the perturbation theory. These
divergences are controlled by an ultraviolet cutoff scale which is
identified as the viscous-dissipation scale $\eta$ acting here as the
renormalization scale.  The summation of the infinite series results
in a factor $(R/\eta)^{2\Delta}$ with some anomalous exponent $\Delta$
which is, generally speaking, of the order of unity. The appearance of
such a factor means that the actual correlation of two $R$-separated
dissipation fields is much larger, when $R$ is much larger than
$\eta$, than the naive prediction of dimensional analysis. The
physical explanation of this renormalization\cite{LL,LPl} is the
effect of the multi-step interaction of two $R$-separated small eddies
of scale $\eta$ with a large eddy of scale $R$ via an infinite set of
eddies of intermediate scales. The net result on the scaling exponent
is that the exponent $\mu$ changes from $8/3$ as expected in the
Kolmogorov theory to $8/3-2\Delta$.

At this point it is important to understand what is the numerical
value of the anomalous exponent $\Delta$. In \cite{LP-2} there was
found an exact sum that forces a relation between the numerical value
of $\Delta$ and the numerical value of the exponent $\zeta_2$ of
$S_2(R)$, $\Delta=2-\zeta_2$.  Such a relation between different
exponents is known in the jargon as a ``scaling relation" or a
``bridge relation". Physically this relation is a consequence of the
existence of a universal nonequilibrium stationary state that supports
an energy flux from large to small scales \cite{LP-2,LP-3}. The
scaling relation for $\Delta$ has far-reaching implications for the
theory of the structure functions. It was explained that with this
value of $\Delta$ the series for the structure functions $S_n(R)$
diverge when the energy-input scale $L$ approaches $\infty$ as powers
of $L$, like $(L/R)^{\delta_n}$. The anomalous exponents $\delta_n$
are the deviations of the exponents of $S_n(R)$ from their Kolmogorov
value. This is a very delicate and important point, and we therefore
expand on it.  Think about the series representation of $S_n(R)$ in
terms of lower order quantities, and imagine that one succeeded to
resum it into an operator equation for $S_n(R)$. Typically such a
resummed equation may look like $[1-\hat O ]S_n(R)]=$RHS, where $\hat
O$ is some integro-differential operator which is not small compared
to unity. If we expand this equation in powers of $\hat O$ around the
RHS we regain the infinite perturbative series that we started with.
However, now we realize that the equation possesses also homogeneous
solutions, solutions of $[1-\hat O ]S_n(R)]=0$ which are inherently
nonperturbative since they can no longer be expanded around a RHS.
These homogeneous solutions may be much larger than the inhomogeneous
perturbative solutions. Of course, homogeneous solutions must be
matched with the boundary conditions at $R=L$, and this is the way
that the energy input scale $L$ appears in the theory.  This is
particularly important when the homogeneous solution diverge in size
when $L \to \infty$ as is indeed the case for the problem at hand.

The next step in the theoretical development is to understand how to
compute the anomalous exponents $\delta_n$. The divergence of the
perturbation theory for $S_n(R)$ with $L\to \infty$ forces us to seek
a nonperturbative handle on the theory. One finds this in the idea
that there exists always a global balance between energy input and
dissipation, which may be turned into a nonperturbative constraint on
each $n$-th order structure function \cite{LP-3}. Using the
Navier-Stokes equations (\ref{NS}) one derives the set of equations of
motion
\begin{equation}
{\partial S_n(R,t) \over \partial t}+D_n(R,t) = \nu J_n(R,t) \ , \label{dSn}
\end{equation}
where $D_n$ and $J_n$ stem from the nonlinear and the viscous terms in
(\ref{NS}) respectively. To understand the physical meaning of this
equation note that $S_2(R)$ is precisely the mean kinetic energy of
motions of size $R$. The term $D_2(R)$ whose meaning is the rate of
energy flux through the scale $R$ is known exactly:
$D_2(R)=dS_3(R)/dR$.  The term $\nu J_2(R)$ is precisely the rate of
energy dissipation due to viscous effects. The higher order equation
for $n>2$ are direct generalizations of this to higher order moments.
In the stationary state the time derivative vanishes and one has the
balance equation $D_n(R) = \nu J_n(R)$. For $n=2$ it reflects the
balance between energy flux and energy dissipation. The evaluation of
$D_n(R)$ for $n>2$ requires dealing with the difficulty (C) of the
nonlocality of the interaction, but it does not pose conceptual
difficulties. It was shown \cite{LP-3} that $D_n(R)$ is of the order
of $dS_{n+1}/dR$. On the other hand, the evaluation of $J_n(R)$ raises
a number of very interesting issues whose resolution lies at the heart
of the universal scaling properties of turbulence.  Presently not all
of these issues have been resolved, and we briefly mention here some
ground on which progress has been made by the present authors.

From the derivation of Eq.(\ref{dSn}) one finds that $J_n(R)$ consists
of a correlation of $\nabla^2 u$ with $n-2$ velocity differences
across a scale $R=|\B.r_1-\B.r_2)|$: $\left \langle\nabla^2 u(\B.r_1)
  [\delta u_{\ell}(\B.r_1,\B.r_2)]^{n-2}\right\rangle$.  The question
is how to evaluate such a quantity in terms of the usual structure
functions $S_n(R)$. Recall that a gradient of a field is the
difference in the field values at two points divided by the separation
when the latter goes to zero. In going to zero one necessarily crosses
the dissipative scale. To understand what happens in this process one
needs first to introduce many-point correlation functions of a product
of $n$ velocity differences:
\begin{equation}
F_n(\B.r_0|\B.r_1,\dots\B.r_n)\equiv\left\langle
\delta u_{\ell}(\B.r_0,\B.r_1)\dots
\delta u_{\ell}(\B.r_0,\B.r_n)\right\rangle \ . \label{Fn}
\end{equation}
Next we need to formulate rules for the evaluation of such correlation
functions of velocity differences when some of the coordinates get
very close to each other. For example, a gradient $\partial/\partial
r_\alpha$ can be formed from the limit $\B.r_1\to \B.r_0$ when we
divide by $r_{1,\alpha}- r_{0,\alpha}$.  These rules are known in the
theoretical jargon as ``fusion rules". The fusion rules for
hydrodynamic turbulence were presented in \cite{LP-f}. They show that
when $p$ coordinates in $F_n$ are separated by a small distance $r$,
and the remaining $n-p$ coordinates are separated by a large distance
$R$, then the scaling dependence on $r$ is like that of $S_p(r)$, i.e.
$r^{\zeta_p}$. This is true until $r$ crosses the dissipative scale.
Assuming that below the viscous-dissipation scale $\eta$ derivatives
exist and the fields are smooth, one can estimate gradients at the end
ofthe smooth range by dividing differences across $\eta$ by $\eta$.
The question is, what is the appropriate cross-over scale to smooth
behaviour?  Is there just one cross-over scale $\eta$, or is there a
multiplicity of such scales, depending on the function one is
studying?  For example, when does the above $n$-point correlator
become differentiable as a function of $r$ when $p$ of its coordinates
approach $\B.r_0$?  Is that typical scale the same as the one
exhibited by $S_p(r)$ itself, or does it depend on $p$ and $n$ and on
the remaining distances of the remaining $n-p$ coordinates that are
still far away from $\B.r_0$ ?

The answer is that there is a multiplicity of cross-over scales. For
the $n$-point correlator discussed above we denote the dissipative
scale as $\eta(p,n,R)$, and it depends on each of its arguments
\cite{LP-f,LP-4}. In particular it depends on the inertial range
variables $R$ and this dependence must be known when one attempts to
determine the scaling exponents $\zeta_n$ of the structure functions.
In brief, this line of thought leads to a set of non-trivial scaling
relations. For example we confirm the phenomenologically conjectured
\cite{Fri} ``bridge relation" $\mu=2-\zeta_6$ (in close agreement with
the experimental values) and predict that although the $\zeta_n$ are
not Kolmogorov, they are nevertheless linear in $n$ for large $n$.

\section {summary}

It appears that there are four conceptual steps in the construction of
a theory of the universal anomalous statistics of turbulence on the
basis of the Navier-Stokes equations. First one needs to take care of
the sweeping interactions that mask the scale invariant theory
\cite{87BL,LP-1}. After doing so the perturbation expansion converges
order by order, and the Kolmogorov scaling of the velocity structure
functions is found as a perturbative solution. Secondly one
understands the appearance of the viscous-dissipation scale $\eta$ as
the natural normalization scale in the theory of the correlation
functions of the gradient fields \cite{LL,LP-2}. This step is similar
to critical phenomena and it leads to a similarly rich theory of
anomalous behaviour of the gradient fields. Only the tip of the
iceberg was considered above. In fact when one considers the
correlations of tensor fields which are constructed from $\partial
u_\alpha/\partial r_\beta$ (rather than the scalar field $\epsilon$)
one finds that every field with a different transformation property
under the rotation of the coordinates has its own independent scaling
exponent which is the analog of $\Delta$ above \cite{96LPP}. The third
step is the understanding of the divergence of the diagrammatic series
for the structure functions as a whole\cite{LP-3}. This sheds light on
the emergence of the energy-input scale $L$ as a normalization length
in the theory of turbulence.  This means that the Kolmogorov basic
assertion that there is no typical scale in the expressions for
statistical quantities involving correlations across a scale $R$ when
$\eta\ll R \ll L$ is doubly wrong. In general both lengths appear in
dimensionless combinations and change the exponents from the
predictions of dimensional analysis.  Examples of correlation
functions in which both normalization scales $L$ and $\eta$ appear
simultaneously were given explicitly \cite{96LPP}. Last but not least
is the formulation of the fusion rules and the exposition of the
multiplicity of the dissipative scales which should eventually result
in a satisfactory description of all the scaling properties
\cite{LP-4}.

The road ahead is not fully charted, but it seems that some of the
conceptual difficulties have been surmounted. We believe that the
crucial building blocks of the theory are now available, and they
begin to delineate the structure of the theory. We hope that the
remaining 4 years of this century will suffice to achieve a proper
understanding of the anomalous scaling exponents in turbulence.
Considerable work, however, is still needed in order to fully clarify
many aspects of the problem, and most of them are as exciting and
important as the scaling properties. There are universal aspects that
go beyond exponents, such as distribution functions and the eddy
viscosity, and there are important non-universal aspects like the role
of inhomogeneities, the effect of boundaries and so on.  Progress on
these issues will bring the theory closer to the concern of the
engineers. The marriage of physics and engineering will be the
challenge of the 21st century.

\acknowledgments Our thinking about these issues were influenced by
discussions with V.  Belinicher, R. Benzi, P. Constantin, G.
Falkovich, U. Frisch, K. Gawedzki, S. Grossmann, L.P. Kadanoff, R.H.
Kraichnan, V.V. Lebedev, M. Nelkin, E. Podivilov, A. Praskovskii, K.R.
Sreenivasan, P. Tabeling, S.  Thomae and V.E Zakharov. We thank them
all. Our work has been supported in part by the Minerva Center for
Nonlinear Physics, by the Minerva Foundation, Munich, Germany, the
German-Israeli Foundation, the US-Israel Binational Science Foundation
and the the Naftali and Anna Backenroth-Bronicki Fund for Research in
Chaos and Complexity.

\end{document}